\documentclass[12pt, onecolumn]{my-article} %
\usepackage{oneinchmargins}
\usepackage{times}
\usepackage{enumerate}
\usepackage{graphicx}
\usepackage{url}
\usepackage[usenames,dvipsnames]{xcolor}
\usepackage[bookmarks=false]{hyperref}

\hypersetup{colorlinks=true,
citecolor=Maroon,
linkcolor=Green,
urlcolor=Maroon}

\usepackage{setspace}
\usepackage{rotating}

\usepackage{floatflt}
\usepackage{alltt}
\usepackage{epstopdf}
\usepackage{amsmath}

\usepackage{fancyvrb}

\usepackage{afterpage}

\begin{document}

\newcommand{\vten}{\vspace{10pt}}
\newcommand{\vfive}{\vspace{5pt}}
\newcommand{\vthree}{\vspace{3pt}}

\newcommand{\vminfive}{\vspace{-5pt}}
\newcommand{\vminten}{\vspace{-10pt}}
\newcommand{\vminfifteen}{\vspace{-15pt}}
\newcommand{\vmintwenty}{\vspace{-20pt}}

\newcommand{\ub}[1]{\underline{{\bf #1}}}
\newcommand{\ts}[1]{{\tt{\small#1}}}
\newcommand{\bquote}{\vspace{-0.25cm} \begin{quote}}
\newcommand{\equote}{\end{quote}\vspace{-0.2cm} }
\def \sec {\S}

\def \yes {$\surd$}

\def \nospace {
  \setlength{\itemsep}{0pt}
  \setlength{\parskip}{0pt}
  \setlength{\parsep}{0pt}
}

\newcommand{\tid}[1]{\textcolor{Purple}{#1}}

\newcommand{\supsection}[1]{\noindent{\Large{\bf #1}}\vten}

\newenvironment{enumerate2}{
  \begin{enumerate}
  \setlength{\itemsep}{1pt}
  \setlength{\parskip}{0pt}
  \setlength{\parsep}{0pt}
}{
  \end{enumerate}
}

\newenvironment{itemize2}{
  \begin{itemize}
 \renewcommand{\labelitemi}{-}
  \setlength{\itemsep}{1pt}
  \setlength{\parskip}{0pt}
  \setlength{\parsep}{0pt}
}{
  \end{itemize}
}

\newcommand{\hsg}[1]{\textcolor{red}{{\small {\bf (HSG: #1)}}}}
\newcommand{\thanh}[1]{\textcolor{red}{{\small {\bf (THANH: #1)}}}}
\newcommand{\achien}[1]{\textcolor{red}{{\small {\bf (ACHIEN: #1)}}}}
\newcommand{\todo}[1]{\textcolor{red}{{\small {\bf (TODO: #1)}}}}
\newcommand{\newtext}[1]{\textcolor{red}{#1}}
\newcommand{\bluetext}[1]{\textcolor{blue}{#1}}

\newcommand{\infokernel}{\mbox{infokernel}}
\newcommand{\unix}{{\sc Unix}}
\newcommand{\dare}{DARE}
\def \fate {\textsc{Fate}}

\def \late {\textsc{Late}}

\def \lights {\textsc{LigHTS}}

\def \nb {\noindent $\bullet$~}
\def \ni {\noindent}

\def \vvvnb {\vfifteen \noindent $\bullet$~}
\def \vvnb {\vten \noindent $\bullet$~}
\def \vnb {\vfive \noindent $\bullet$~}

\def \vvn {\vten \noindent}
\def \vn {\vfive \noindent}
\def \nb {\noindent $\bullet$~}
\def \ni {\noindent}


\newcommand{\hypo}[1]{
\begin{quote}
\stepcounter{HYPO}{\bf Hypothesis \arabic{HYPO}:} 
{\em #1}
\end{quote}
}

\newcommand{\taskformat}[2]{#1\textsc{#2}}

\newcommand{\task}[3]{
\begin{quote}
\phantomsection
\hypertarget{task#1#2}{}
{\bf Task \taskformat{#1}{#2}:} 
{\em #3}
\end{quote}
}

\newcommand{\tasklink}[2]{\hyperlink{task#1#2}{\taskformat{#1}{#2}}}

\newcounter{HYPO}
\newcounter{TASK}

\newcommand{\rs}{{ResearchStaff$_1$}}
\newcommand{\pd}{{\bf Postdoc$_1$}}
\newcommand{\raOne}{{\bf RA$_1$}}
\newcommand{\raTwo}{{\bf RA$_2$}}
\newcommand{\ndv}{{\bf NDV}}
\newcommand{\ug}{{\bf Undergrad$_1$}}


\newcommand{\sssubsection}[1]{\vten\textbf{\large{\textsc{#1}}}}

\newcommand{\emptypage}{
\newpage
\thispagestyle{empty}
(empty page)
}

\newcommand{\myrotate}[1]{\begin{rotate}{90} {\bf #1} \end{rotate}}

\newcommand{\mycaption}[3]{
\caption{
\label{#1}
{\bf #2. } 
{\em \small #3}
}}

\newcommand{\eg}{\textit{e.g.}}
\newcommand{\ie}{\textit{i.e.}}
\newcommand{\etal}{\textit{et al.}}
\newcommand{\etc}{etc.}

\newcounter{Xcounter}
\newcommand{\xxxreset}{\setcounter{Xcounter}{1}}
\newcommand{\xxx}{
\textcolor{red}{
\textbf{XXX$_{\arabic{Xcounter}}$}\stepcounter{Xcounter}}}


\title{\textsf{\textbf{Impact of Limpware on HDFS:\\ A Probabilistic Estimation}}}

\author{\textsf{Thanh Do$^{\dag}$ and Haryadi S. Gunawi}
}

\date{
\begin{tabular}{ccc}
\textsf{$^{\dag}$ University of Wisconsin-Madison}
&  & 
\textsf{University of Chicago} 
\end{tabular}
}

\maketitle

\begin{abstract}

{\em 	With the advent of cloud computing, thousands of machines
	  are connected and managed collectively.  This era is confronted
	  with a new challenge: performance variability, primarily caused
	  by large-scale management issues such as hardware failures,
	  software bugs, and configuration mistakes.  
		In our previous work~\cite{DoGunawi13-LimpingHotCloud,
		Do+13-Limplock}
	  we highlighted one overlooked cause: limpware -- hardware
	  whose performance degrades significantly compared to its
	  specification. We showed that limpware can cause
		severe impact in current scale-out systems.
		In this report, we quantify how often these scenarios happen
		in Hadoop Distributed File System.
}

\end{abstract}

\section{Introduction} 
\label{sec:introduction}

In our latest work~\cite{DoGunawi13-LimpingHotCloud, Do+13-Limplock}, 
we highlight one overlooked cause of
performance variability: limpware - hardware whose performance
degrades significantly compared to its specification.  The
growing complexity of technology scaling, manufacturing, design
logic, usage, and operating environment increases the occurrence
of limpware. We believe this trend will continue, and the
concept of performance perfect hardware no longer holds.
We have collected reports and anecdotes on cases of limpware.  We
find that disk bandwidth can drop by 80\%, network throughput by
two orders of magnitude, and processor speed by
25\%. Interestingly, such degraded behavior is exhibited by both
commodity as well as enterprise hardware.
Our work shows that although today’s scale-out systems employ redundancies,
they are not capable of making limpware ``fail in place''.
Impact of limpware cascades, leading to 
degraded operation (\eg, a write can degrade to 1KB without trigger
a failover), nodes and cluster (\eg, 
a node or the whole cluster are unable to perform certain task).

In this report, we calculate how often these degraded scenarios happen
in HDFS~\cite{HDFSArchitecture}. Although, HDFS employs redundancies for
fault-tolerance, its protocols are susceptible to 
limpware~\cite{DoGunawi13-LimpingHotCloud, Do+13-Limplock}.
We specifically look at three protocols
(\ie, read, write, and regeneration) and quantify the probability that these
protocols experience degraded condition.
We further verify our calculation by simulation.
Our results show that probabilities of these scenarios are alarmingly high
in small and medium (\eg, 30-node) clusters. However, 
these probabilities reduce significantly when size of cluster increases, 
as ``Scale can be your friend''~\cite{Ongaro+11-RAMCloud}.

This report is structured as follows.
 We highlight an overview of HDFS in Section~\ref{hdfs},
present the probability derivation in Section~\ref{freq}, and conclude.

\section{HDFS Overivew} 
\label{hdfs}

We now briefly describe the architecture and main operations of 
HDFS~\cite{HDFSArchitecture}.
HDFS has a dedicated master, the \emph{namenode}, and multiple workers
called \emph{datanodes}. The namenode is responsible for
file-system metadata operations, which are handled by a fixed-size 
thread pool with 10 handlers by default.
The namenode stores all metadata, including namespace 
structure and block locations, in memory for fast operations.



While the namenode serves metadata operations, 
the datanodes serve read and write requests.
For fault tolerance, data blocks are replicated across datanodes.
A new data block is written through a pipeline 
of three different nodes by default. Therefore, each data block
typically has three identical replicas. On read, HDFS tries to serve the 
request a replica that is closest to the reader.


Since a data block can be under-replicated due to many reasons
such as disk and machine failures, the namenode ensures that each block has
the intended number of replicas by sending commands to datanodes, asking them
to regenerate certain blocks. 
Block regeneration also happens when a datanode is decommissioned;
all of its blocks are regenerated before it leaves.
Each datanode allows maximumly two threads
serving regeneration request at a time so that regeneration does 
not affect foreground workload. 


\section{Probability Derivation}
\label{freq}
In this section, we first show examples of limpware causing
negative impact on three protocols of HDFS: read, write, and regeneration.
We then calculate how often such scenarios happen for each protocol.

\subsection{Impact of Limpware}
\def \figwidth {1.0in} 

\begin{figure*}
	\centerline{
	\includegraphics[width=0.5in]{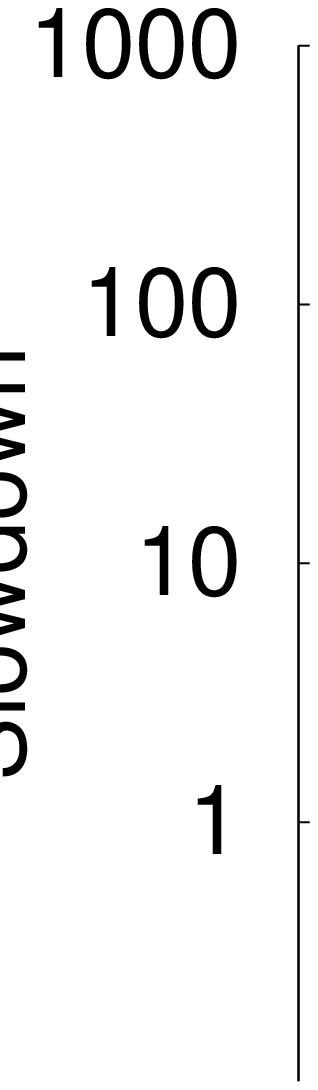}
	\includegraphics[width=\figwidth]{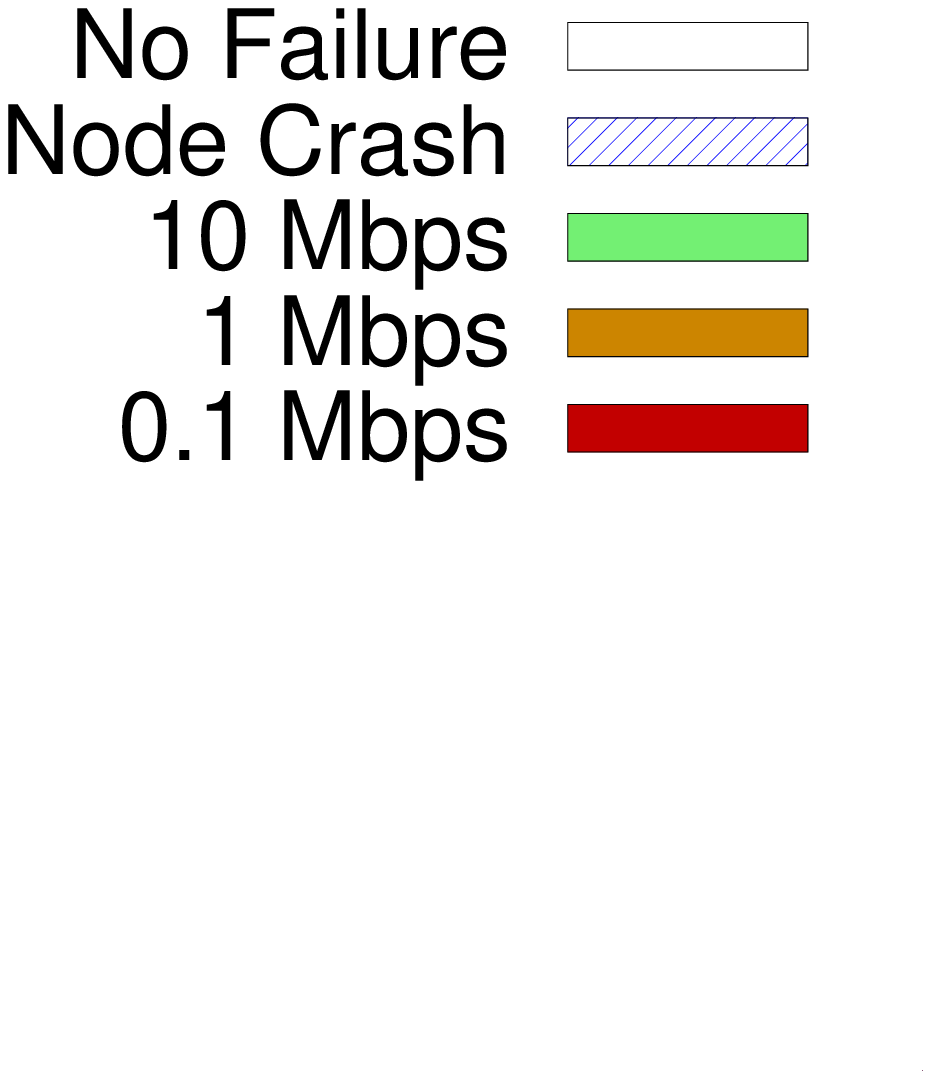}
	\includegraphics[width=\figwidth]{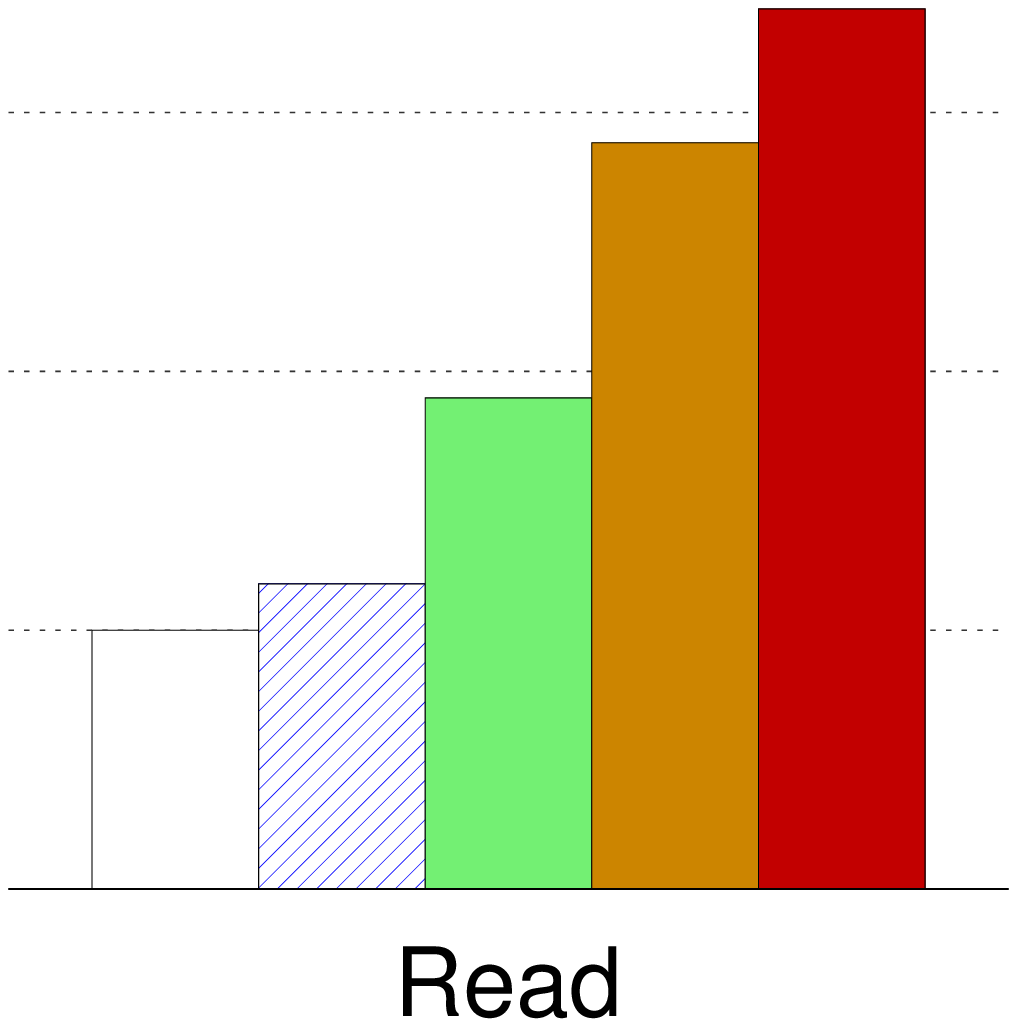}
	\includegraphics[width=\figwidth]{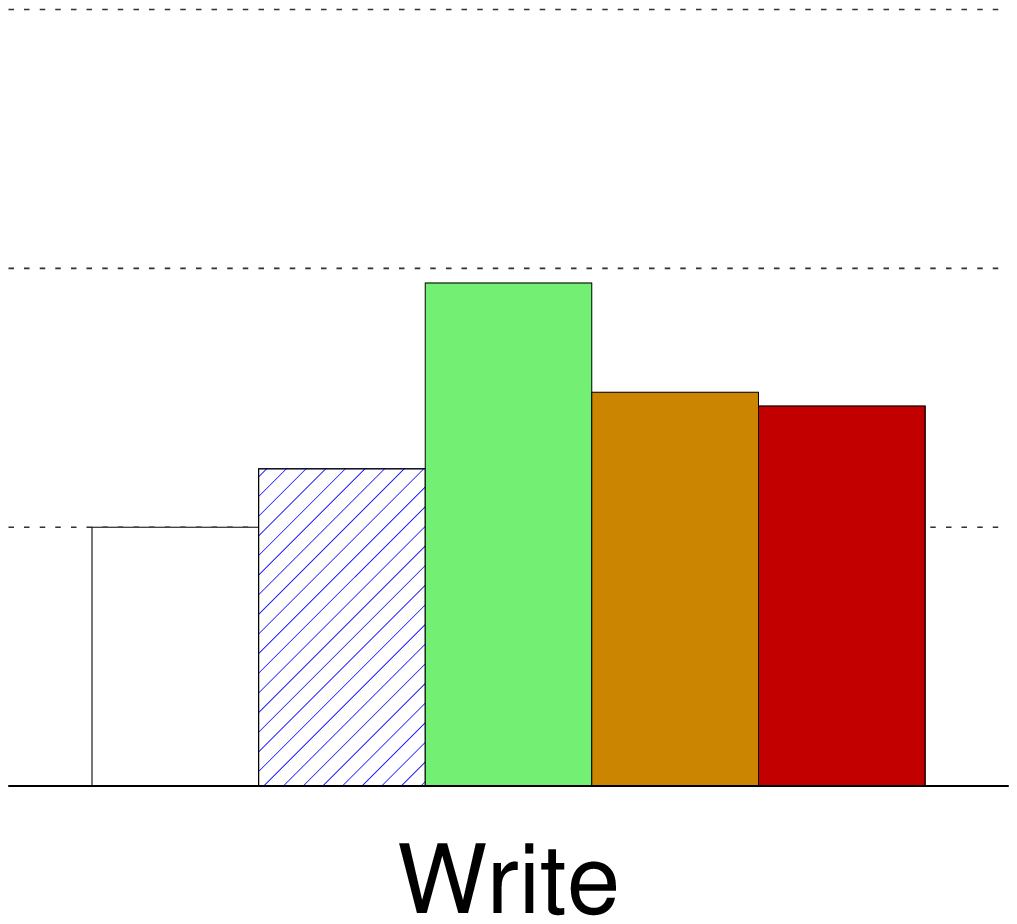}
	\includegraphics[width=\figwidth]{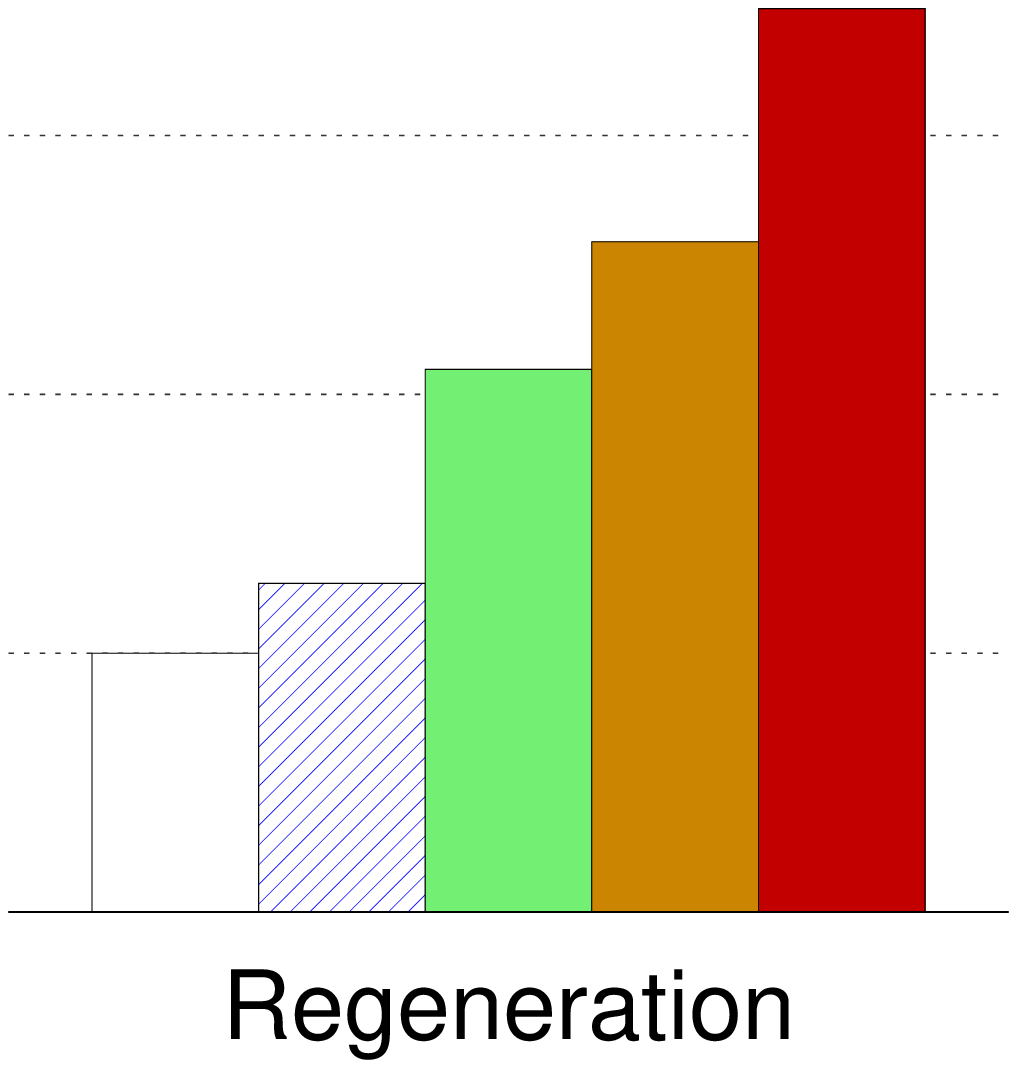}
	}
	\mycaption{fig-impact}{Impact of limpware on HDFS}{The figures
	show the impact of slow network card on three HDFS
	protocols: read, write and regeneration.}
\end{figure*}

Our previous work~\cite{DoGunawi13-LimpingHotCloud} shows that
HDFS is limpware intolerant. Here, we show examples
of HDFS protocols suffering from negative impact of slow network card (NIC).
Specifically, we run workloads that exercise three HDFS protocols
(read, write, and regeneration), inject slowdown to NIC of a node
in the cluster, and measure the resulting execution time. Figure~\ref{fig-impact}
shows the results. The normal bandwidth for the network is 100Mbps. 
We slow down the NIC to 10, 1, and 0.1Mbps in each experiment.
We inject crash to evaluate HDFS fail-stop failure tolerance.
In all experiments, HDFS is not able to detect a slow NIC,
hence does not trigger a failover. As a result, total execution time
in case of slow NIC is orders of magnitude higher than in normal scenario.

These results confirm that HDFS protocols are not able to tolerate
limpware. We next quantify how often such negative impacts happen
for each protocol, given the cluster's size, number of data blocks it 
manages, and number of user requests.

\subsection{Degraded Read}
\label{freq-read}

\vvnb {\bf Definition.} 
Consider an $n$-node cluster which has one slow node $L$ and $n-1$ good
nodes. A user request reads data from one out of three copies 
(assuming 3-way replication) of certain block $B$.  
Each copy has an equal chance to be chosen.
We define a \emph{degraded read} to be a 
read request that reads data the slow node $L$.

\begin{figure*}[t]
	\centerline{
	\includegraphics[width=2.5in]{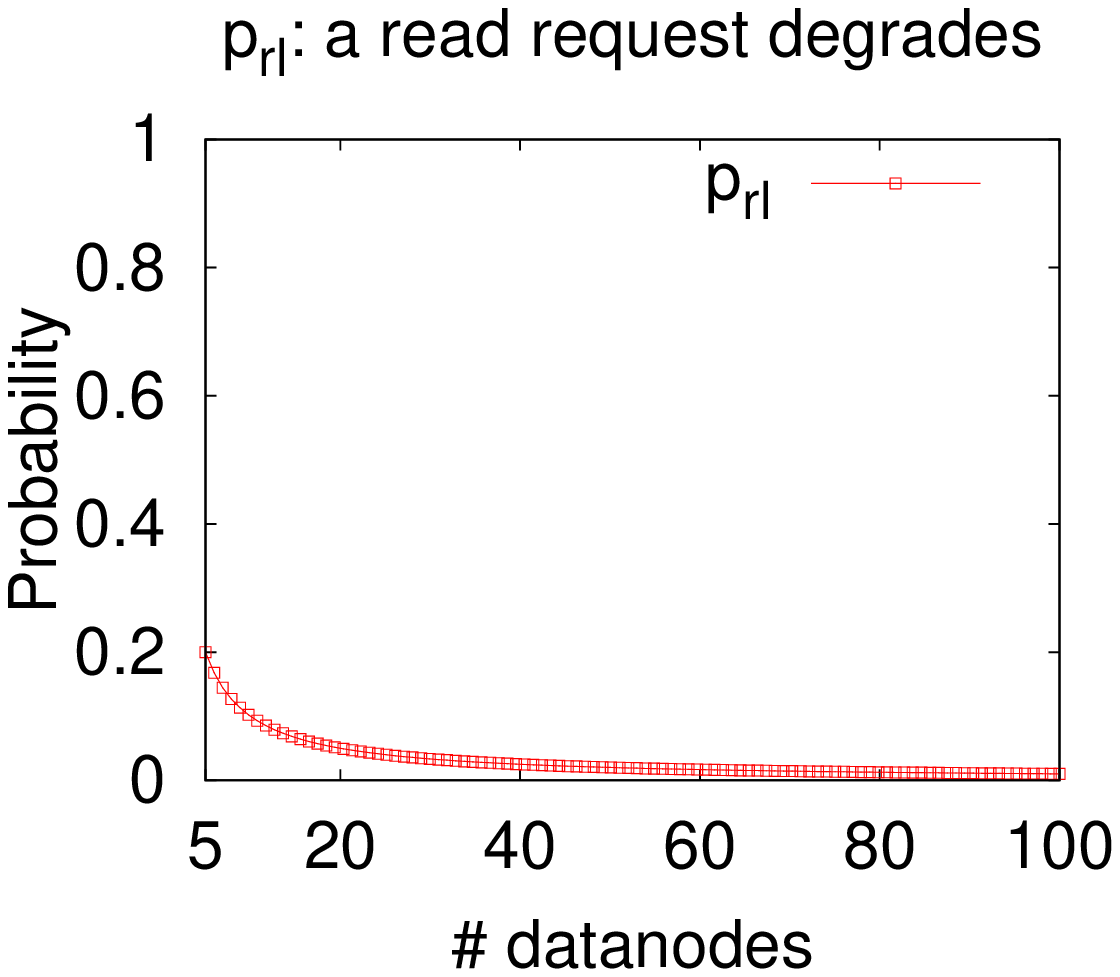}
  \includegraphics[width=2.5in]{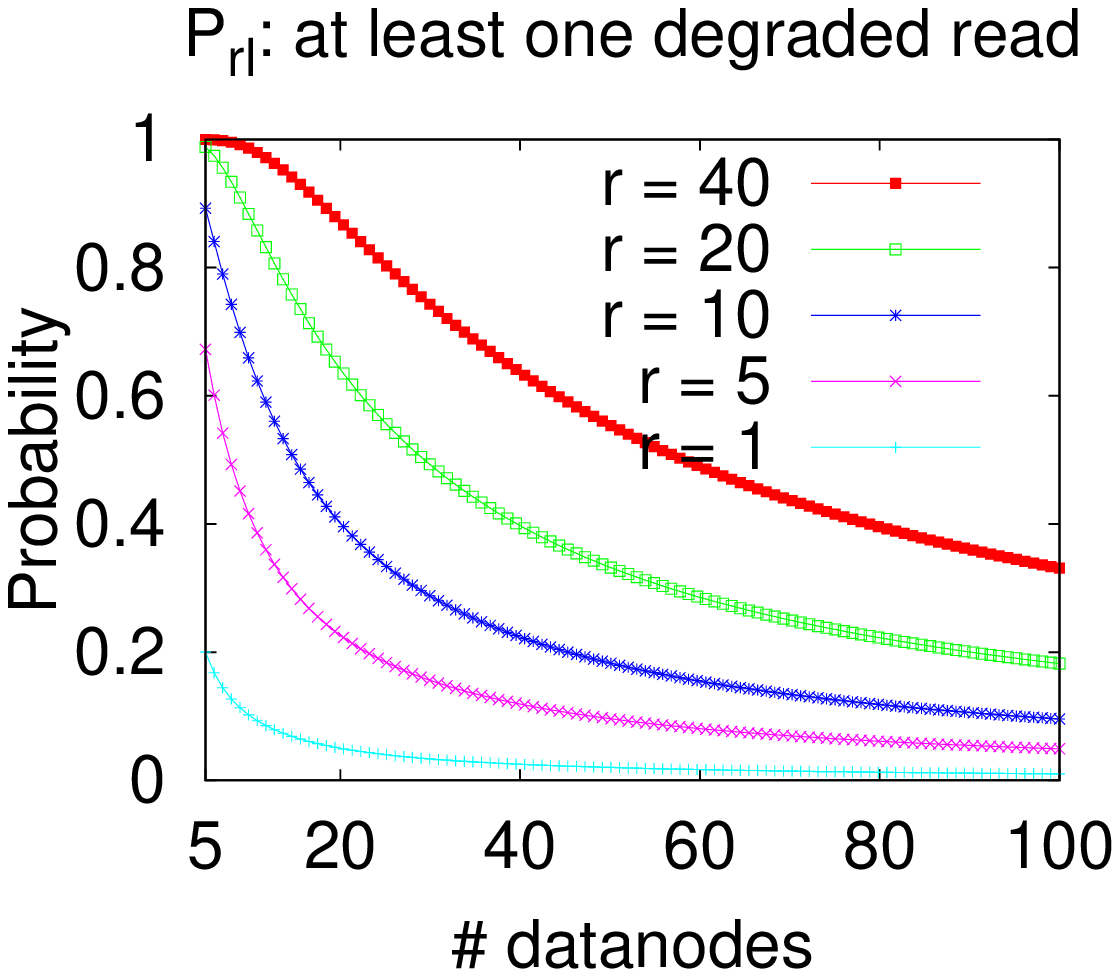}
	}
	\mycaption{fig-read}{Degraded Read Probability}{}
\end{figure*}

\vvnb {\bf Derivation.} 
We now derive the probability of a degraded read.
There are two conditions for a read of block $B$ to degrade. 
First, $L$ must contain one copy of $B$, and second,
the copy in $L$ is chosen for reading.

Let's derive the probability for the first condition.
There are ${n \choose 3}$ ways to choose 3 out of $n$ nodes; there are
${n-1 \choose 3}$ ways to choose 3 out of $n-1$ \emph{good} nodes.
Therefore, the number of ways to choose 3 nodes, one of which is $L$,
out of $n$ nodes is ${n \choose 3} - {n-1 \choose 3}$. The probability
for L to contain one copy of $B$ is:
\begin{equation}
\begin{split}
P(L~contains~one~copy~of~B) = \frac{{n \choose 3} - {n-1 \choose 3}}{{n \choose 3}} = \frac{3}{n}
\end{split}
\end{equation}

Since there are three copies of $B$, the probability for the copy in 
slow node $L$ to be chosen for reading is $\frac{1}{3}$.
As a result, the probability for a read to degrade is:
\begin{equation}
	P(a~read~to~degrade) = p_{rl} = \frac{3}{n} \times \frac{1}{3} = \frac{1}{n}
\end{equation}

Let $r$ be the number of read requests of a user during a certain
operation period (\eg, a day). We now derive the probability 
that the user has at least one degraded read. The probability 
for a read \emph{not} to degrade is $1 - p_{rl}$. The probability for
\emph{all} $r$ requests \emph{not} to degrade is $(1 - p_{rl})^r$.
As a result, the probability for a user to experience at least one degraded
read is:
\begin{equation}
	P(user~has~at~least~one~degraded~read) = P_{rl} = 1 - (1 - p_{rl})^r = 1 - (1-\frac{1}{n})^r
\end{equation}

\vvnb {\bf Result.} Figure~\ref{fig-read} plots probabilities for a request to
degrade ($p_{rl}$) and for a user to experience at least 
one degraded read ($P_{rl}$). As cluster size increases, these probabilities
decrease since there are more healthy nodes.

\subsection{Degraded Write}

\begin{figure*}
	\centerline{
	\includegraphics[width=2.5in]{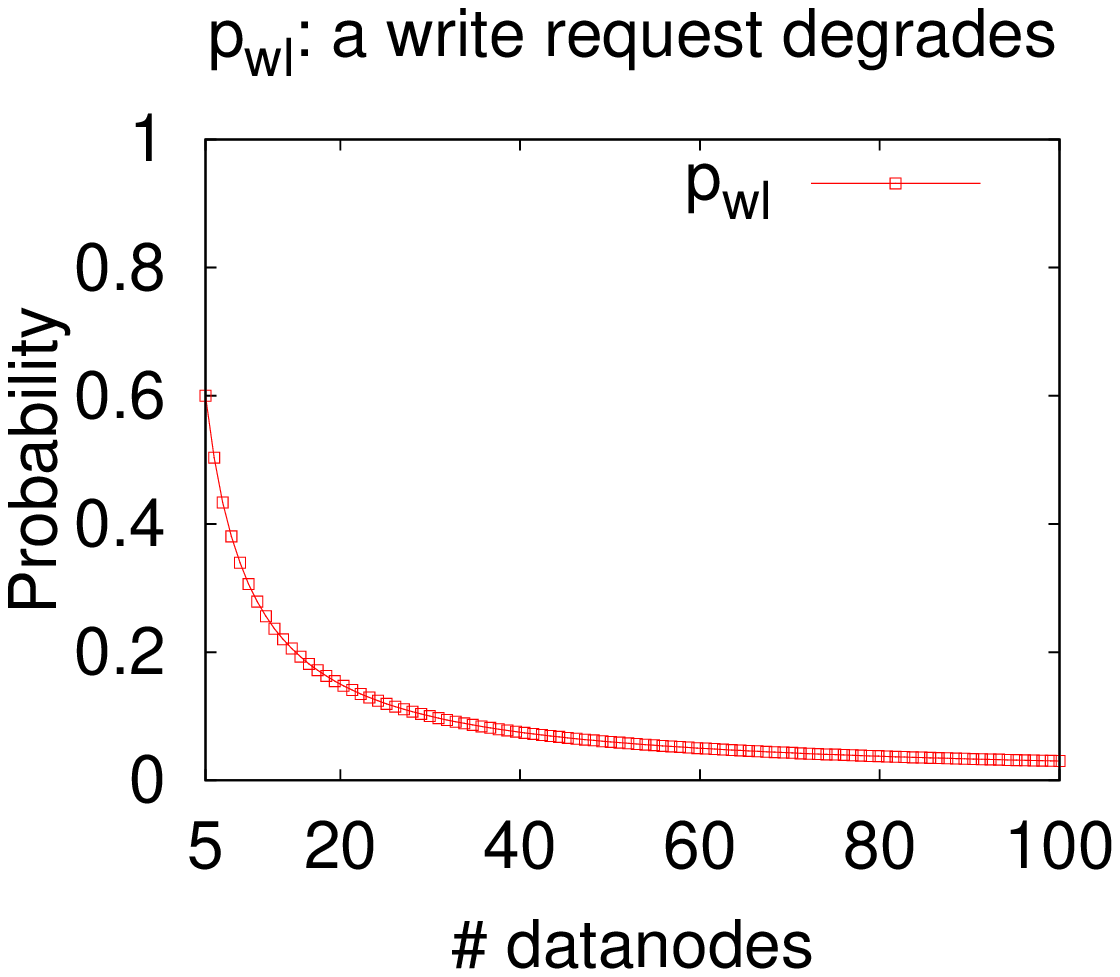}
		\includegraphics[width=2.5in]{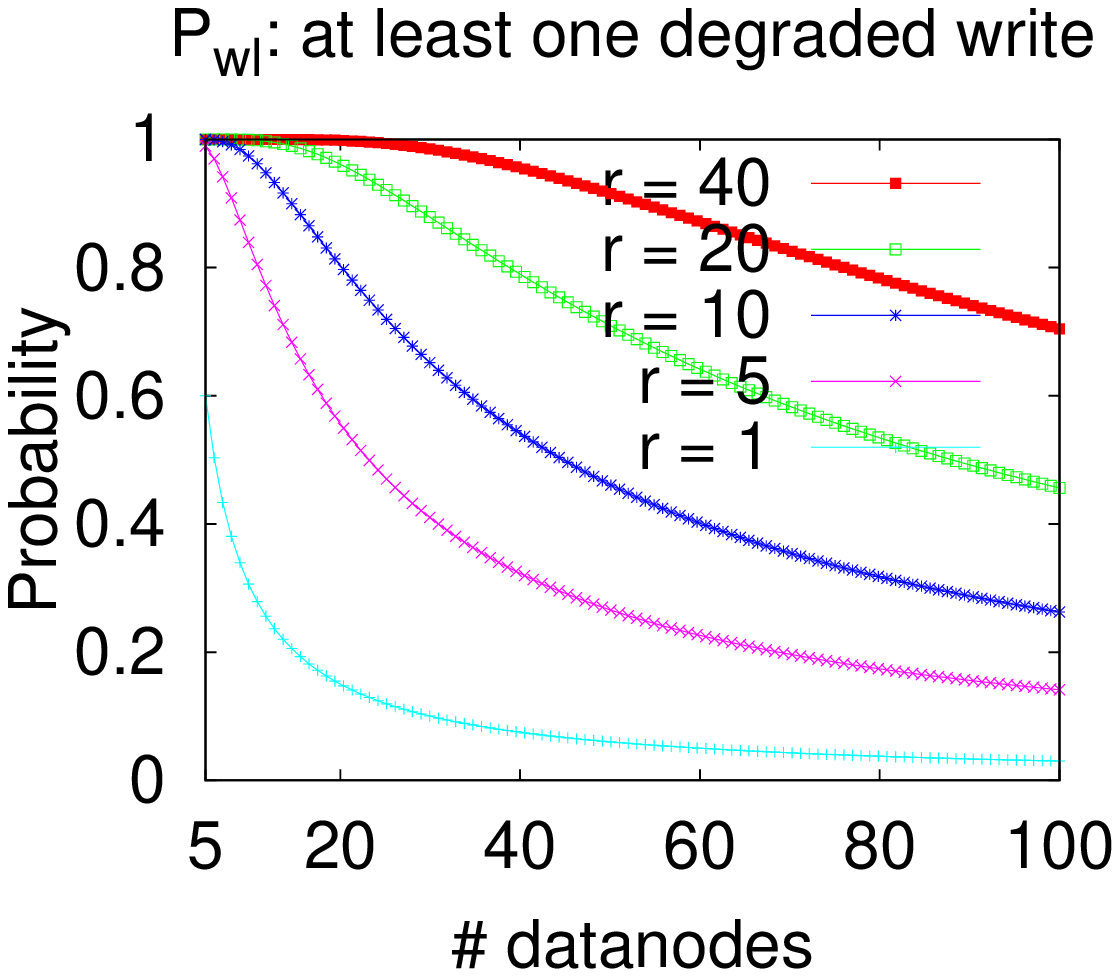}
	}
  \mycaption{fig-write}{Degraded Write Probability}{}
\end{figure*}

\vvnb {\bf Definition.} 
Consider an $n$-node cluster which has one slow node $L$ and $n-1$ good
nodes. A user write request requires HDFS to allocate 3 nodes 
to write to (assuming 3-way replication). Each node has an equal chance to 
be chosen in a write pipeline.
We define a \emph{degraded write} to be a write request whose 
pipeline contains $L$.

\vvnb {\bf Derivation.} 
We now derive the probability for write to be slow. It is the probability for
$L$ to be chosen as one of the nodes in the 3-node write pipeline.
We follow the similar derivation as in Section~\ref{freq-read}.
There are ${n \choose 3}$ ways to choose 3 out of $n$ nodes; there are
${n-1 \choose 3}$ ways to choose 3 out of $n-1$ \emph{good} nodes.
Therefore, the number of ways to choose 3 nodes, one of which is $L$,
out of $n$ nodes is ${n \choose 3} - {n-1 \choose 3}$.
Thus, the probability for a write to be degraded is:

\begin{equation}
P(a~write~to~degrade)=p_{wl} = \frac{{n \choose 3} - {n-1 \choose 3}}{{n \choose 3}} = \frac{3}{n}
\end{equation}

Let $r$ be the total number of requests that a user has during a certain
working period (e.g., a day). We now derive the formula for the probability 
that the user experience  at least one slow write, $P_{wl}$.
The probability for a write \emph{not} to be slow is $1 - p_{rl}$. 
The probability for the user does \emph{not} have any slow write 
equals the probability that \emph{all} $r$ write requests are \emph{not} degraded,
which is $(1 - p_{wl})^r$. Therefore, the probability for the user experiences
at least one degraded write is:
\begin{equation}
P(user~has~at~least~one~degraded~write)=P_{wl} = 1 - (1 - p_{wl})^r = 1 - (1-\frac{3}{n})^r
\end{equation}

\vvnb {\bf Result.} Figure~\ref{fig-write} plots the probabilities
for degraded write as function of cluster size and number of user requests.
These probabilities are significant larger than those for degraded read,
because each write has to be written to a 3-node pipeline.
Even in a cluster of 50 nodes, a user is likely to experience one slow
write on every 40 requests.

\subsection{Degraded Regeneration}

\begin{figure*}[t]
	\centerline{
	\includegraphics[width=6.6in]{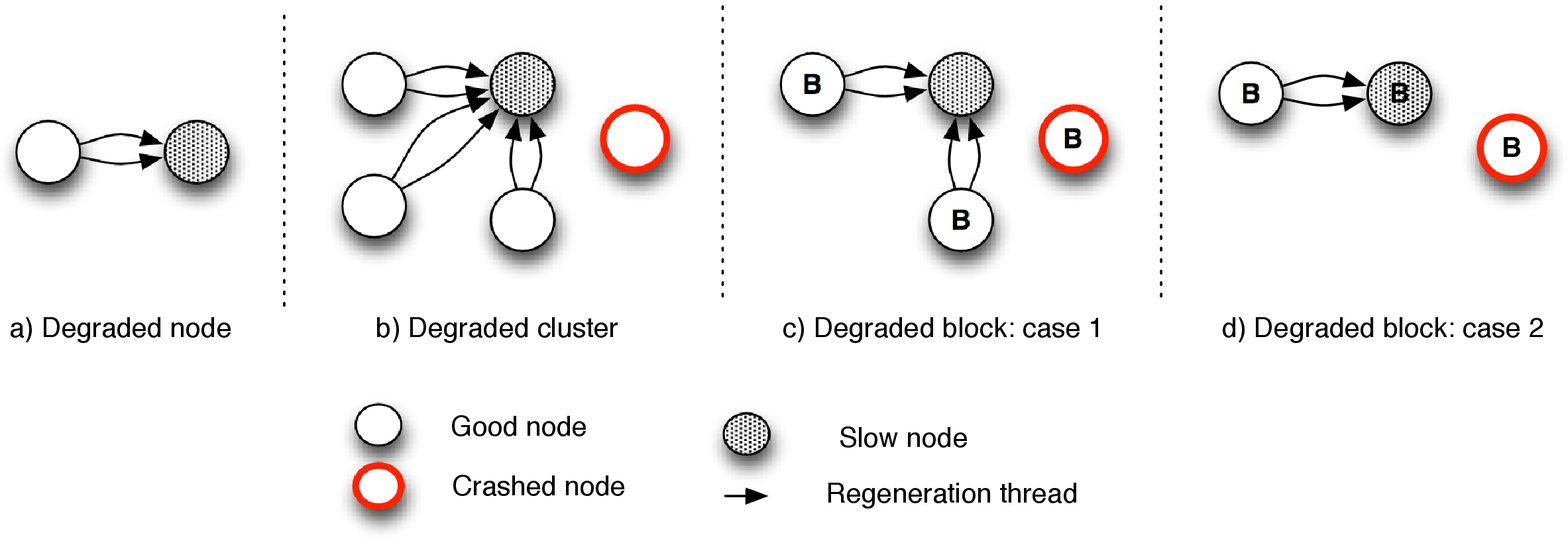}
	}
	\mycaption{fig-regeneration-il}{Degraded Regeneration}{The figures show
	different scenarios of degraded regeneration. ``B'' label inside a circle
	represents that a copy of block $B$ is located in the node.
	Arrow represents a regeneration	thread, which may be copying a block
	other than $B$.}
\end{figure*}

\subsubsection{Definitions}
\label{sec-def}
Consider an HDFS cluster consisting of $n$ datanodes, one of which 
is slow (node $L$). Let $C$ be a node that crashes; 
there are $n-1$ surviving nodes including the slow one.
Let $G$ be the set of good nodes (nodes are neither slow nor crashed);
there are $n-2$ good nodes.

Let $b$ be the total number of blocks in node $C$. 
When node $C$ crashes, HDFS triggers regeneration workload
to regenerate those lost blocks.
On average, each surviving node has to
replicate  $m=\frac{b}{n-1}$ blocks to other live nodes.

\begin{equation}
m=\frac{b}{n-1}
\end{equation}

For each lost block, the master chooses a source and a destination datanode. 
The source is chosen from live nodes that still carry the block. 
The destination node is chosen using the write allocation policy 
(that uses randomness). A source datanode can only run two regeneration
threads at a time.

\vvnb {\bf Degraded node.} 
Consider a good node $X, X \in G$.
When both regeneration threads of $X$ send blocks to a slow node $L$,
the node is not available for new regeneration tasks
until the two threads finish (which could take a long time). 
We define this situation a \emph{degraded node} during regeneration process.
The situation is illustrated in Figure~\ref{fig-regeneration-il}a.

\vvnb {\bf Degraded cluster.} 
When {\em all} good nodes are degraded, as illustrated in
Figure~\ref{fig-regeneration-il}b, the
whole system is unable to start \emph{any} regeneration task. We
define this scenario a \emph{degraded cluster}.
Formally, the cluster degrades during regeneration 
when $\forall X \in G, X$ degrades.

\vvnb {\bf Degraded block.} The system may not be able to regenerate
block $B$ for a long time. This can happen in two cases, which
are illustrated in Figures~\ref{fig-regeneration-il}c and 
\ref{fig-regeneration-il}d.
First, all remaining copies of $B$ are in degraded nodes
(Figures~\ref{fig-regeneration-il}c),
and second, one copy of B is in a degraded node, the other is in slow node $L$
(Figures~\ref{fig-regeneration-il}d).
Note that these cases are mutually exclusive and
in the illustration, replication threads are copying different blocks
other than $B$.

\subsubsection{Derivation}

We now derive the probabilities for degraded node, cluster, and block
scenarios. To facilitate our calculation, we first derive the 
probability that node $L$ is destination of a copy task.

\vvnb {\bf $L$ is destination for a copy task.} 
Consider a scenario where good node $X$ $(X \in G)$ copies one of its blocks
(\eg, block $B$) to another node. Let $p$ be the probability 
that $L$ is selected as destination. For this to happen,
there are two conditions: first, $L$ does not have a copy of $B$ \emph{and} 
second, the master chooses $L$ to be the destination.

We now derive the probability of the first condition.
Since $X$ and $C$ both contain a copy of $B$, the probability for
$L$ to also contain $B$ is $\frac{1}{n-2}$. Therefore, the probability
for a copy of $B$ not in $L$ is $1-\frac{1}{n-2}=\frac{n-3}{n-2}$.

\begin{equation}
P(copy ~ of ~ block ~ B ~ in ~ L) = \frac{1}{n-2}
\end{equation}

\begin{equation}
P(copy ~ of ~ block ~ B ~ not ~ in ~ L) = 1- \frac{1}{n-2}=\frac{n-3}{n-2}
\end{equation}

We calculate the probability that the master chooses $L$ as destination,
given that $L$ does not contain $B$.
Note that to the master can only choose one from $n-3$ nodes that do not have
a copy of block $B$. Thus, given $L$ not storing $B$,
the probability for $L$ to be the destination is $\frac{1}{n-3}$.
As a result, the probability for $X$ to copy block $B$ to $L$ is:

\begin{equation}
p = P(L ~ is ~ destination ~ of ~ a ~ copy ~ task)=\frac{n-3}{n-2}\times\frac{1}{n-3}=\frac{1}{n-2}
\end{equation}

\vvnb {\bf Degraded node probability.} 
Let $P_{nl}$ be the probability for node $X, X \in G$ degrades
during regeneration process.
We assume the time to copy a block between two good nodes
is inconsiderable compared to the time to copy a block between a good and 
a slow node $L$. As a result, $P_{nl}$ is the probability that $X$ copies
\emph{at least} two blocks to $L$, out of $m$ blocks it has
to regenerate.

Since the probability for X {\em not} to copy \emph{any} blocks to L
(out of $m$ blocks) is $(1-p)^m$ and the probability for X to copy 
\emph{exactly} one block to L (again, out of $m$ blocks) is
${m \choose 1} \times p \times (1-p)^{m-1}$, we have:

\begin{equation}
\begin{split}
P(a ~ node ~ degrades) = P_{nl} \\
= 1 - (1-p)^m - {m \choose 1} \times p \times (1-p)^{m-1} \\
= 1 - (1-\frac{1}{n-2})^{\frac{b}{n-1}} - \frac{b}{(n-1) \times (n-2)} \times (1-\frac{1}{n-2})^{\frac{b-n+1}{n-1}}
\end{split}
\end{equation}

\vvnb {\bf Degraded cluster probability.} 
Let $P_{cl}$ be the probability for the whole cluster to degrade during
regeneration. This scenario happens when all good nodes degrade.
Therefore:
\begin{equation}
	P(the ~ cluster ~ degrades) = P_{cl} = {P_{nl}}^{n-2}
\end{equation}

\vvnb {\bf Degraded block probability.} 
Let $p_{bl}$ be the probability for a block $B$ to be degraded.
There are two mutually exclusive cases for this scenario
(Figure~\ref{fig-regeneration-il}c and Figure~\ref{fig-regeneration-il}d).
Let the probabilities of these cases are $p_{bl_{1}}$ and $p_{bl_{2}}$,
respectively. Because they are mutually exclusive, we have:

\begin{equation}
p_{bl} = p_{bl_{1}} + p_{bl_{2}}
\end{equation}

We now calculate probability of the first case, $p_{bl_{1}}$, 
the case where all of block $B$'s remaining copies
are stored in degraded nodes (Figure~\ref{fig-regeneration-il}c).
Let $i$ be the number of good but degraded nodes.  
The probability to have \emph{exactly} $i$ good but degraded nodes is
\begin{equation}  
\begin{split}
P(having ~ i ~ degraded ~ nodes) = \\ 
p_{nl}(i) = {n-2 \choose i} \times {P_{nl}}^i \times (1-P_{nl})^{n-2-i}
\end{split}
\end{equation}

For this first case to happen, there are two conditions:
(1) $i \geq 2$; and
(2) two copies of block $B$ are stored among those $i$ nodes.
There are ${n-1 \choose 2}$ ways to place two copies of $B$ among $n-1$ nodes
(excluding the crashed one which must contain $B$).
There are ${i \choose 2}$ ways to place two copies of $B$ among $i$ degraded
nodes. Therefore, the probability for two copies of $B$ be in two (out of $i$)
degraded nodes is 
$\frac{{i \choose 2}}{{n-1 \choose 2}}$.
As a result:

\begin{equation}
p_{bl_{1}}(i) = p_{nl}(i) \times \frac{{i \choose 2}}{{n-1 \choose 2}}, 2 \leq i \leq n-2
\end{equation}

To calculate the exact value of $p_{bl_{1}}$, we must consider all possible
values of $i$. Because $i$ can vary from $2$ to $n-2$, the final equation
for the probability of the first case (Figure~\ref{fig-regeneration-il}c) is:
\begin{equation}
p_{bl_{1}} = \sum_{i=2}^{n-2} p_{nl}(i) \times \frac{{i \choose 2}}{{n-1 \choose 2}}
\end{equation}

Now, let's calculate, $p_{bl_{2}}$, the probability for the second case
(Figure~\ref{fig-regeneration-il}d),
which happens when: (1) $i \geq 1$; and (2) one remaining copy of $B$ is 
in $L$, and the other is in one (out of $i$) good but degraded node.
Again, the probability to have \emph{exactly} $i$ good but degraded nodes
is $p_{nl}(i)$. There are ${i \choose 1} = i$ ways to place two copies of $B$,
one of which in $L$ and the other one in degraded node. 
Therefore, the probability for two copies of $B$ be in this situation is
$\frac{i}{{n-1 \choose 2}}$. As a result:
\begin{equation}
p_{bl_{2}}(i) = p_{nl}(i) \times \frac{i}{{n-1 \choose 2}}, 1 \leq i \leq n-2
\end{equation}

Since $i$ can vary from $1$ to $n-2$ in the second case, we have:
\begin{equation}
	p_{bl_{2}} = \sum_{i=1}^{n-2} p_{nl}(i) \times \frac{i}{{n-1 \choose 2}}
\end{equation}

Since two cases for a block to degrade are mutually exclusive, the degraded
block probability is:

\begin{equation}
	\begin{split}
P(a ~ degraded ~ block) =	p_{bl} = p_{bl_{1}} + p_{bl_{2}} \\
= \sum_{i=2}^{n-2} p_{nl}(i) \times \frac{{i \choose 2}}{{n-1 \choose 2}} + \sum_{i=1}^{n-2} p_{nl}(i) \times \frac{i}{{n-1 \choose 2}}
\end{split}
\end{equation}

We are now able to calculate the probability for the scenario where at least
one block degrades during regeneration process. The probability for a block B
not to degrades is $1-p_{bl}$. The probability of having \emph{zero} degraded
block is $(1-p_{bl})^b$. Therefore,
the probability of having at least one degraded block:

\begin{equation}
\begin{split}
P(at~least~one~degraded~block) = \\ 
P_{bl} = 1 - (1-p_{bl})^b
\end{split}
\end{equation}

\subsubsection{Results}
\def \figwidth {2.0in} 

\begin{figure*}
	\centerline{
	\includegraphics[width=\figwidth]{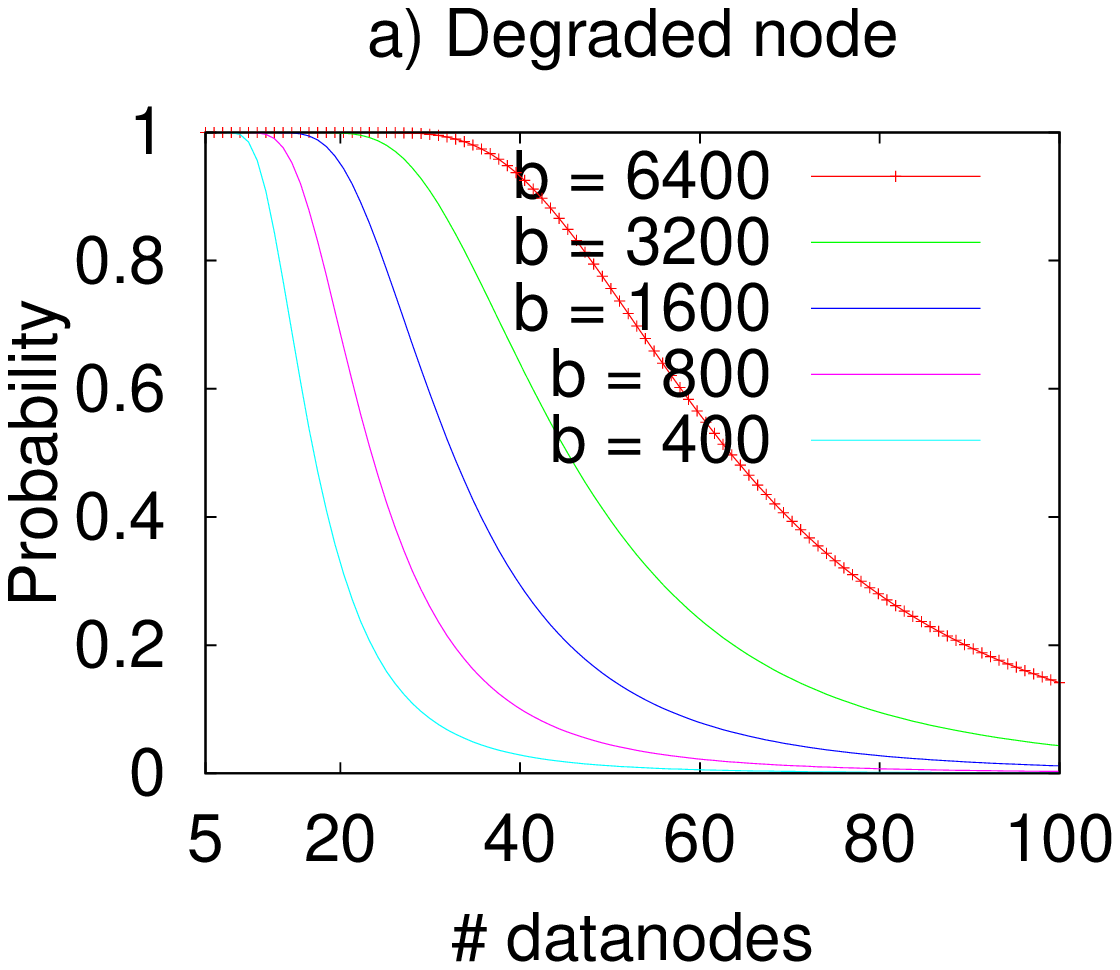}
	\includegraphics[width=\figwidth]{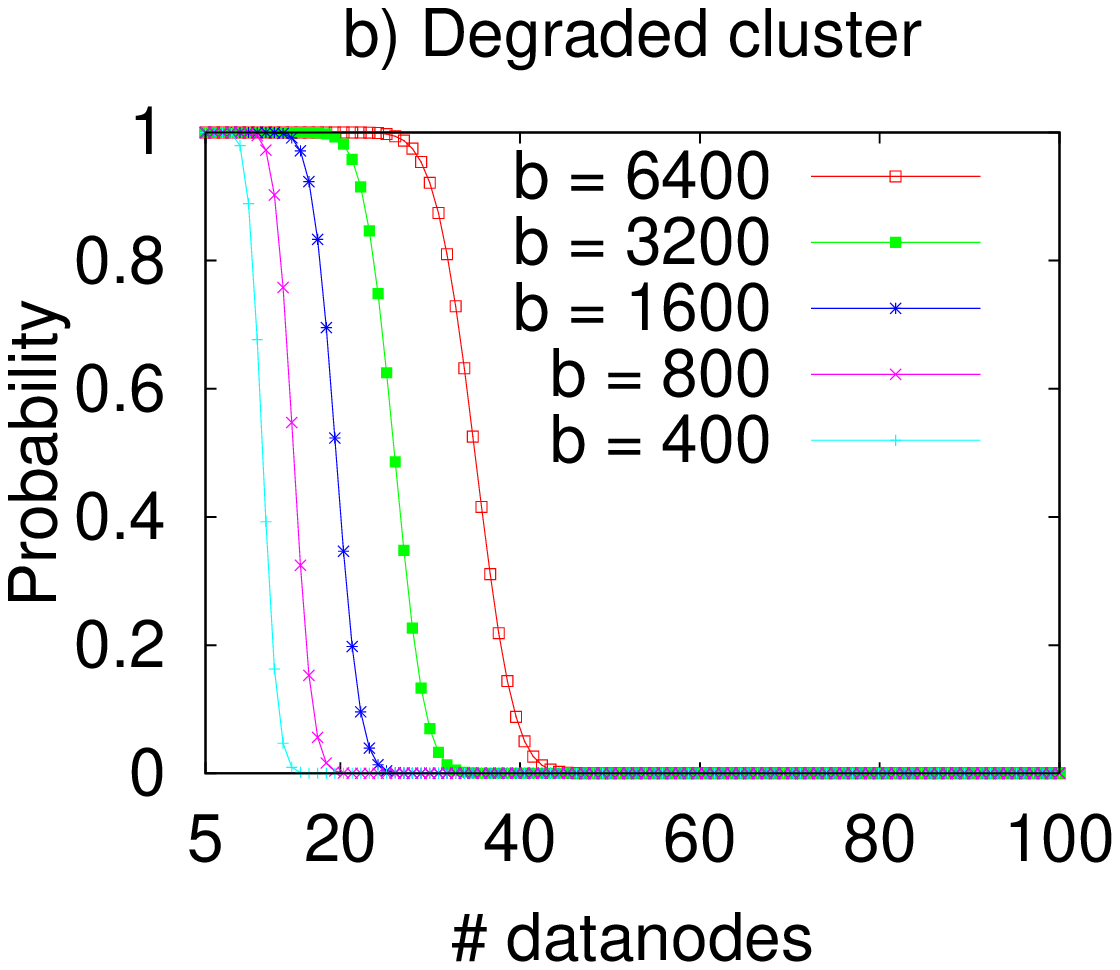}
	\includegraphics[width=\figwidth]{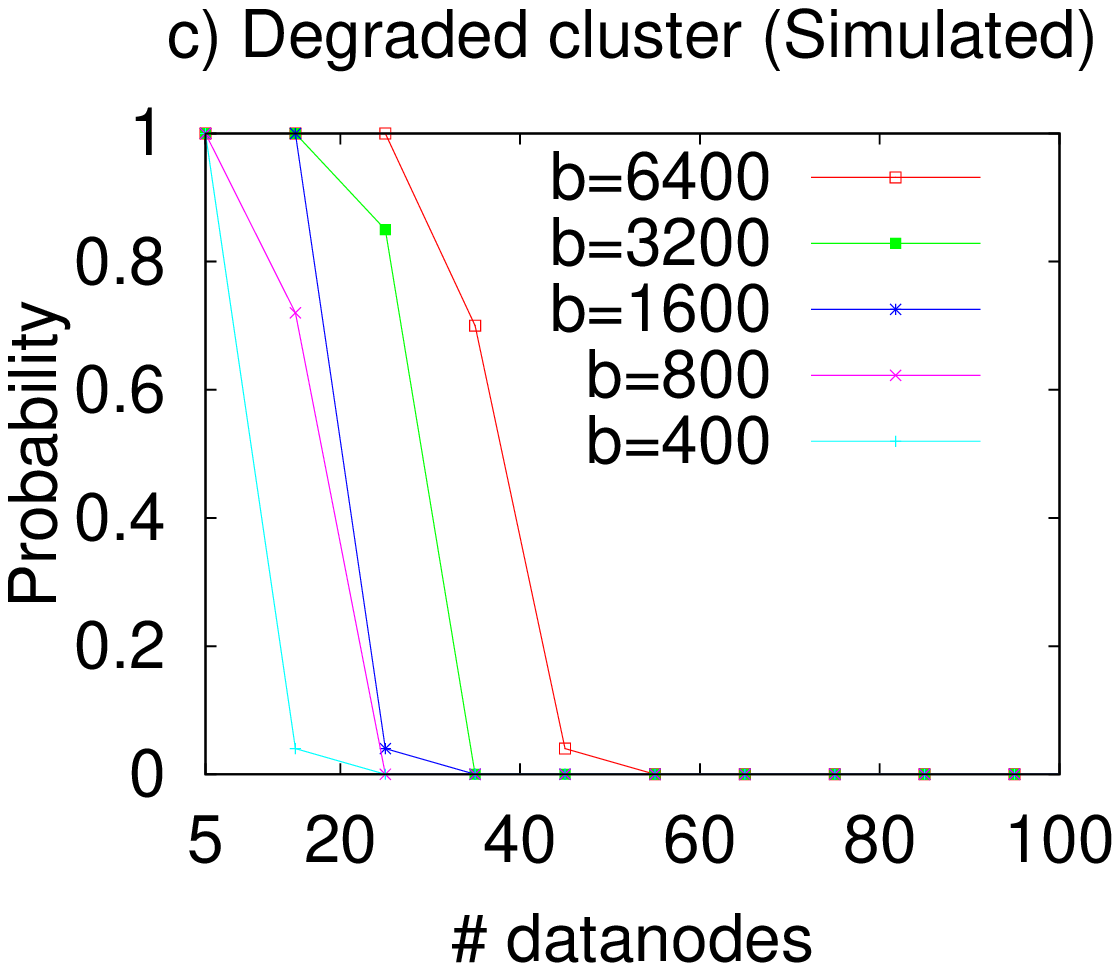}
	}
	\mycaption{fig-node-cluster}{Degraded node and cluster probabilities}{}
\end{figure*}  

\def \figwidth {2.0in} 

\begin{figure*}
	\centerline{
	\includegraphics[width=\figwidth]{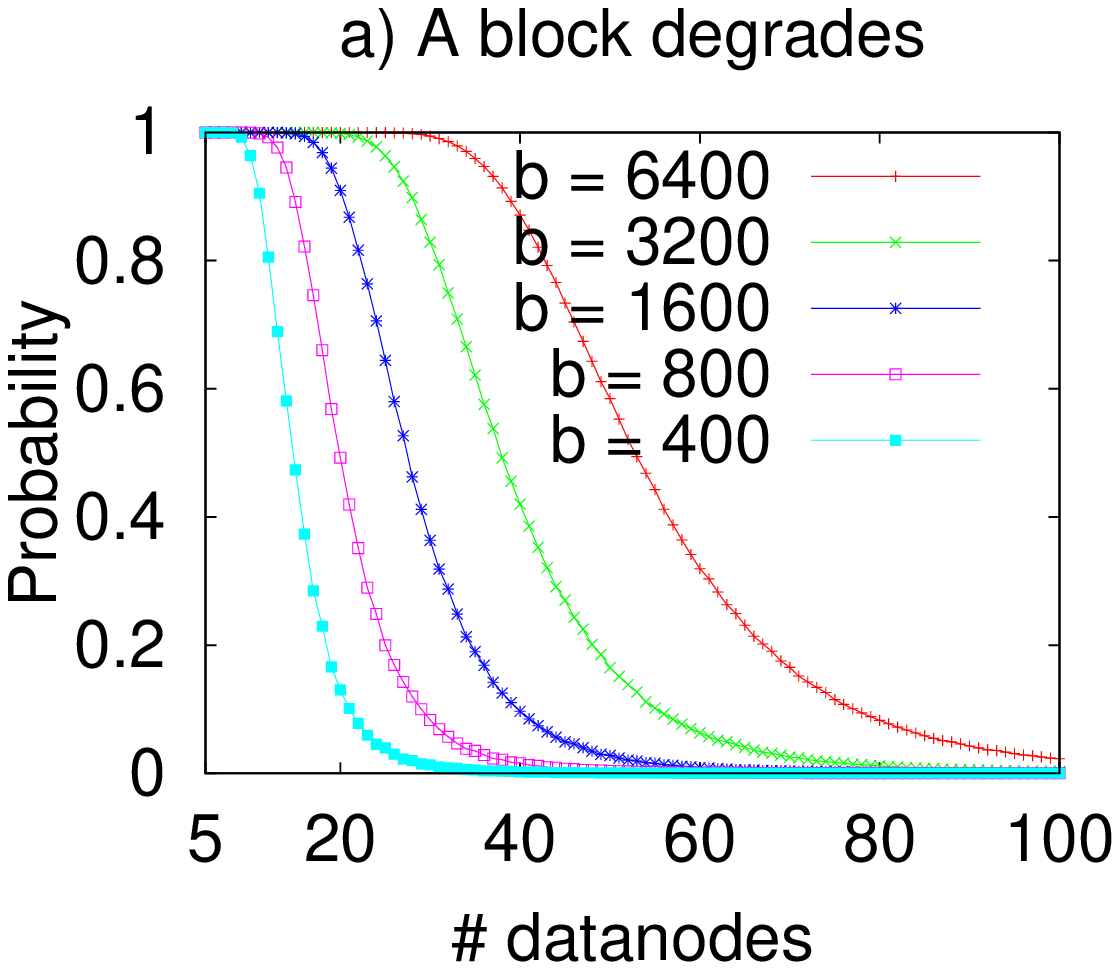}
	\includegraphics[width=\figwidth]{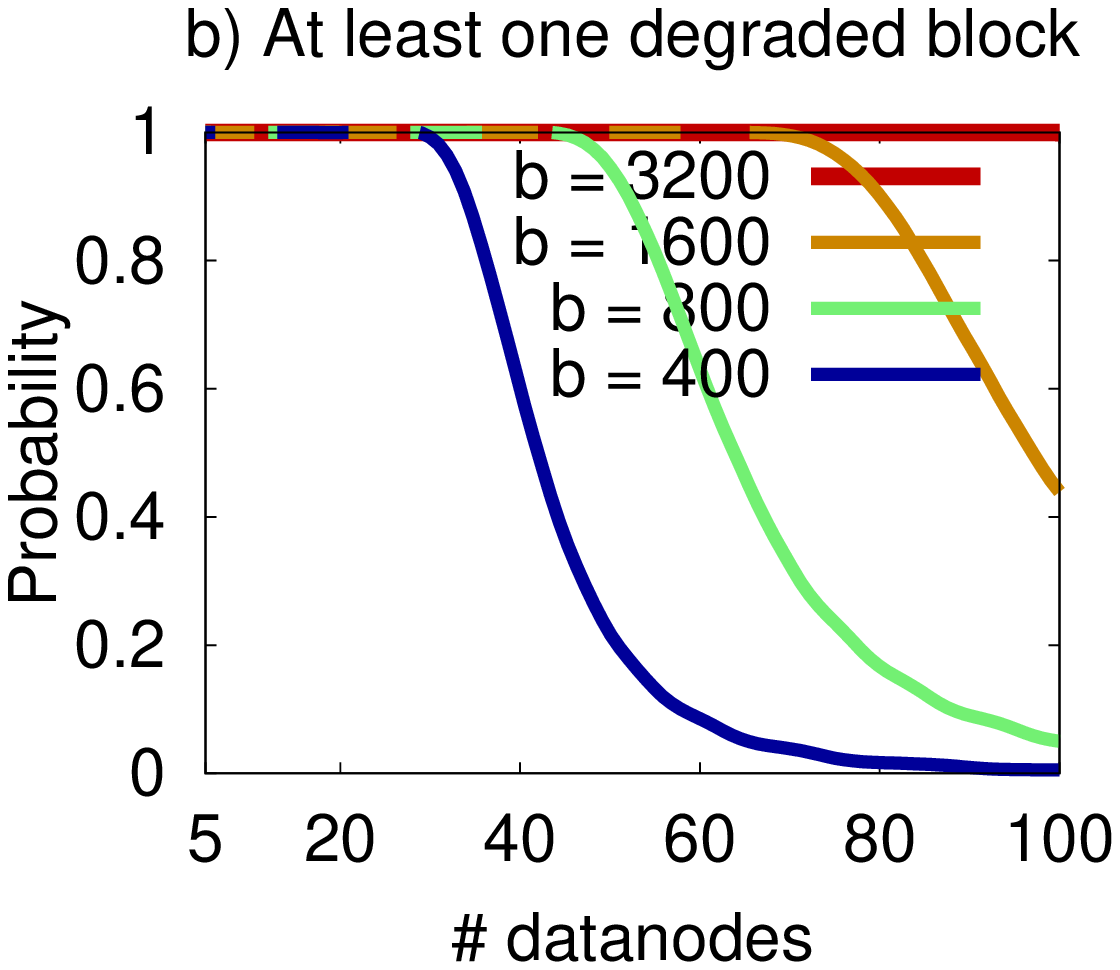}
	\includegraphics[width=\figwidth]{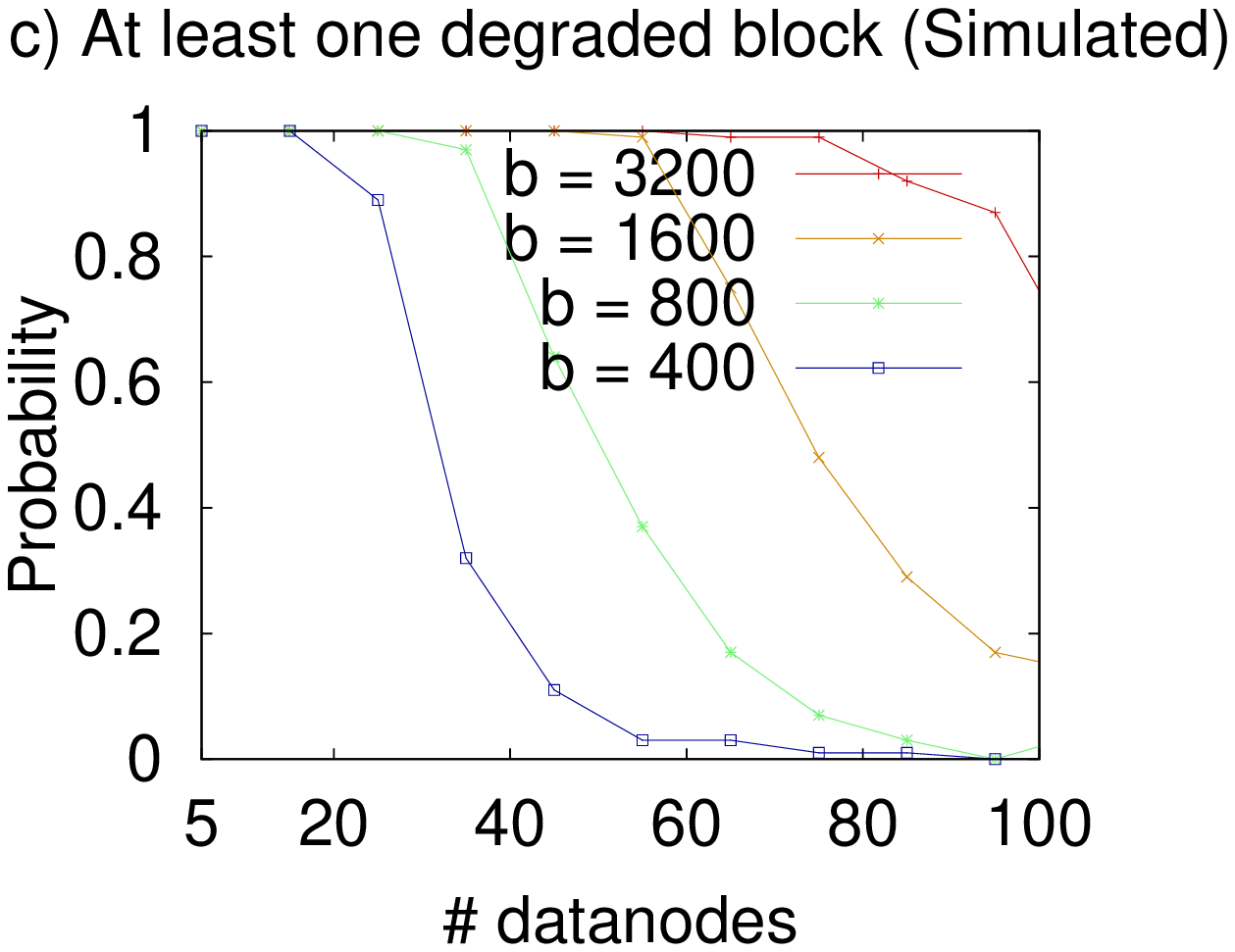}
	}
	\mycaption{fig-block}{Degraded block probabilities}{}
\end{figure*}                     

To be more confident with our calculation, we simulate HDFS regeneration
protocol and run regeneration workload. We vary the number of nodes
in the cluster and the number of lost blocks.
We run each configuration (with different cluster size and number of lost
blocks) 100 times, and measures the probability of degraded block and 
degraded cluster.

Figures~\ref{fig-node-cluster} and ~\ref{fig-block} show both our calculation
and simulation results. 
Degraded-node and degraded-cluster probabilities are  relatively high for a small
to medium (e.g., 30-node) cluster.
Degraded block probability is alarmingly high: even in a 100-node cluster,
a dead 20\%-full 1TB node
(that can store 3200 blocks) will lead to at least one degraded block.
Simulation results are similar to our calculation.

\section{Conclusion} 
\label{sec:conclusion}

Limpware without doubt is a destructive failure mode, yet we
show that HDFS fail to properly handle limpware.
We present a probabilistic estimation of how often such negative impact
of limpware happens to three important HDFS protocols: read,
write, and regeneration. Our estimation shows that impact of 
limpware is significant, even a medium sized cluster of 30-40 nodes.






{\small

   \bibliographystyle{plain}

   \bibliography{local-socc}

}

\end{document}